\documentclass[intlimits,twoside,a4paper]{article}

\usepackage{amsmath,amssymb}

\usepackage[T2A]{fontenc}
\usepackage[cp1251]{inputenc}
\usepackage{cmpj2}

\articletype{Rapid Communication}

\issue{2012}{15}{2}{24001}

\doinumber{10.5488/CMP.15.24001}

\title[The method of collective variables: a link with the DFT]%
{The method of collective variables: a link with the density functional theory}

\author{O. Patsahan, I. Mryglod}

\authorcopyright{O. Patsahan, I. Mryglod, 2012}

 \address{Institute for Condensed Matter Physics of the National Academy of Sciences of Ukraine,\\
 1 Svientsistski Str., 79011 Lviv, Ukraine}
\date{Received March 14, 2012, in final form May 3, 2012}
\begin{document}

\maketitle

\begin{abstract}
Recently, based on the method of collective variables the
statistical field theory for  multicomponent inhomogeneous systems
was formulated [O. Patsahan, I. Mryglod, J.-M. Caillol, Journal of
Physical Studies, 2007, {\textbf {11}}, 133]. In this letter we
establish a link between this approach  and the classical density
functional theory for inhomogeneous fluids.
 \pacs 05.20.Jj, 02.30.Sa
\end{abstract}

The  powerful tools for the study of equilibrium and non-equilibrium
properties of many-particle interacting systems are those based on
the functional methods. In many cases the partition function of such
systems  can be re-expressed as a functional integral after
performing the Hubbard-Stratonovich (HS) transformation
\cite{stratonovich,hubbard}, a simple device proposed in the 50ies.
Nearly at the same time another method, the method of collective
variables (CVs), that allows one in an explicit way to derive a
functional representation for many-particle interacting systems, was
developed \cite{zubar,jukh}. The method, proposed initially in the
1950s \cite{bohm,zubar,jukh} for the description of the classical
charged many particle systems and developed later for the needs of
the phase transition theory
\cite{yuk,yuk_nuovo_cimento,yuk2,patyuk3,patyuk4}, was in fact one
of the first successful attempts to attack the problems of
statistical physics using the functional integral representation.
The CV method is based on: (i) the concept of collective coordinates
being appropriate for the physics of the system considered (see, for
instance, \cite{Yuk-Hol}) and (ii) the functional integral identity
\begin{eqnarray}
\label{a}
\exp \left(F[\hat{\rho}]
\right) = \int \mathcal{D} \rho \;
 \delta_{\mathcal{F}}\left[ \rho -\widehat{\rho} \right]
 \exp \left(
F[\rho]\right)
\end{eqnarray}
valid for classical systems and permitting to derive an exact functional representation for the
configurational Boltzmann factor. Being  applied to the fluids, the CV method uses the
idea of the reference system (RS), one of the basic ideas in the liquid state theory \cite{hansen_mcdonald}.

Recently, the rigorous scalar field KSSHE
(Kac-Siegert-Stratonovich-Hubbard-Edwards) theory
\cite{Cai-Mol,Cai-JSP}, which uses the HS transformation, was
developed to describe the phase equilibria in simple and ionic
fluids. As was shown \cite{caillol_patsahan_mryglod,
patsahan_mryglod_CM}, both theories (KSSHE and CVs)  are in close
relation.

Another valuable theoretical approach, which has been extensively
employed to study the structural and thermodynamic properties of
inhomogeneous systems is the classical density-functional theory
(DFT) (for an overview we refer to \cite{Evans, Singh}). The central
quantity of DFT is the Helmholtz excess free energy expressed as a
functional of the single-particle density. There are few systems for
which this functional is known exactly. Thus,  successful
application of   DFT critically depends on judicious choice of an
appropriate Helmholtz energy functional suitable for the system
under investigation \cite{hansen_mcdonald}. The main goal of this
letter is to constitute a link between the CVs based theory and the
classical DFT.

Let us consider the  general case of a classical $m$-component
system consisting of $N$ particles among which there exist $N_{1}$
particles of species $1$, $N_{2}$ particles of species $2$, \ldots
and  $N_{m}$ particles of species $m$.The potential energy of the
system is assumed to be of the form
\begin{equation}
{\cal U}_{N_{1}\ldots N_{m}}=\frac{1}{2}\sum_{\alpha,\beta}^{m}\sum_{i\neq j}^{N_{\alpha},N_{\beta}}U_{\alpha\beta}({\mathbf r}_{i},{\mathbf r}_{j})+\sum_{\alpha=1}^{m}\sum_{i=1}^{N_{\alpha}}
\psi_{\alpha}({\mathbf r}_{i}),
\label{2.1}
\end{equation}
where $U_{\alpha\beta}({\mathbf r}_{i},{\mathbf r}_{j})$ denotes the interaction potential of two particles and the second term is
the potential energy due to external forces.

The pair interaction potential $U_{\alpha\beta}({\mathbf r}_{i},{\mathbf r}_{j})$  can be considered as the sum
\begin{equation}
U_{\alpha\beta}({\mathbf r}_{i},{\mathbf r}_{j})=v_{\alpha\beta}^{0}({\mathbf r}_{i},{\mathbf r}_{j})+w_{\alpha\beta}({\mathbf r}_{i},{\mathbf r}_{j}),
\label{split}
\end{equation}
where $v_{\alpha\beta}^{0}({\mathbf r}_{i},{\mathbf r}_{j})$ is a
potential of a short-range repulsion which in general describes the
mutual impenetrability of the particles, while
$w_{\alpha\beta}({\mathbf r}_{i},{\mathbf r}_{j})$,  on the
contrary, mainly describes  the behaviour at moderate and large
distances. The system with the interaction potential
$v_{\alpha\beta}^{0}({\mathbf r}_{i},{\mathbf r}_{j})$ can be
regarded as the reference system (RS). The fluid of  hard spheres is
most frequently used as the RS in the liquid state theory since its
thermodynamic and structural properties are well known.

Introducing the microscopic density of the $\alpha$th species in a
given configuration
\begin{equation*}
\widehat\rho_{\alpha}({\mathbf r})=\displaystyle\sum_{i=1}^{N_{\alpha}}\delta({\mathbf r}-{\mathbf r}_{i})
\end{equation*}
we can present the  grand canonical partition function of the system as follows \cite{Patsahan_Mryglod_Caillol-JPS}:
\begin{eqnarray}
\Xi[\{\nu_{\alpha}\}]&=&\sum_{N_{1}\geqslant
0}\frac{1}{N_{1}!}\sum_{N_{2}\geqslant
0}\frac{1}{N_{2}!}\ldots\sum_{N_{m}\geqslant  0}\frac{1}{N_{m}!} \int({\rm
d}\Gamma)\exp\left[-\beta{\cal V}^{RS}_{N_{1}\ldots N_{m}}
-\frac{\beta}{2}\langle\widehat\rho_{\alpha}\vert
w_{\alpha\beta}\vert\widehat\rho_{\beta}\rangle
+\langle\bar\nu_{\alpha}\vert\widehat\rho_{\alpha}\rangle\right]
\label{2.4}
\end{eqnarray}
with  $(\rm
d\Gamma)=\prod_{\alpha}{\rm d}\Gamma_{N_{\alpha}}$, ${\rm
d}\Gamma_{N_{\alpha}}={\rm d}{\mathbf r}_{1}^{\alpha}{\rm d}{\mathbf
r}_{2}^{\alpha}\ldots{\rm d}{\mathbf r}_{N_{\alpha}}^{\alpha}$, being the element of the
configurational space of $N$ particles.
In the right hand side of equation (\ref{2.4})  Dirac's  brackets notations
\begin{eqnarray*}
\sum_{\alpha\beta}\int {\rm d}{\mathbf r}{\rm d}{\mathbf r'}\;
\;\widehat\rho_{\alpha}({\mathbf r})w_{\alpha\beta}({\mathbf
r},{\mathbf r'})\widehat\rho_{\beta}({\mathbf r'})
&=&\langle\widehat\rho_{\alpha}\vert
w_{\alpha\beta}\vert\widehat\rho_{\beta}\rangle,
\\
\sum_{\alpha}\int {\rm d}{\mathbf r} \;\psi_{\alpha}({\mathbf
r})\widehat\rho_{\alpha}({\mathbf
r})&=&\langle\psi_{\alpha}\vert\widehat\rho_{\alpha}\rangle
\end{eqnarray*}
are introduced and  summation over repeated indices is meant.
In  (\ref{2.4}),  ${\cal V}^{RS}_{N_{1}\ldots N_{m}}$
denotes the contribution from a $m$-component RS,
\begin{equation}
\overline\nu_{\alpha}({\mathbf
r})=\nu_{\alpha}+\nu_{\alpha}^{\mathrm{S}}-\beta\psi_{\alpha}({\mathbf r})
\label{nu_bar}
\end{equation}
is the local chemical potential of the $\alpha$th
species, $\nu_{\alpha}=\beta\mu_{\alpha}-3\ln\Lambda_{\alpha}$,
$\Lambda_{\alpha}$ is
the de Broglie thermal wavelength and $\nu_{\alpha}^{\mathrm{S}}$ is
the self-energy of the $\alpha$th species $\nu_{\alpha}^{\mathrm{S}}=\beta
w_{\alpha\alpha}({\mathbf r},{\mathbf r})/2$. For a given volume $V$,
$\Xi[\{\nu_{\alpha}\}]$ is a function of the temperature $T$ and a
log-convex functional of the local chemical potentials
$\nu_{\alpha}({\mathbf r})$.

Using (\ref{a}) we can present the Boltzmann factor which does not include the RS interaction in the form
\begin{eqnarray}
\label{b}
\exp \left(
\frac{1}{2}\left\langle \widehat{\rho}_{\alpha}\vert w_{\alpha\beta} \vert \widehat{\rho}_{\beta}\right\rangle
\right)& =& \int \mathcal{D} \rho  \mathcal{D} \omega \;
\exp \left(  \frac{1}{2}\left\langle \rho_{\alpha} \vert w_{\alpha\beta} \vert \rho_{\beta} \right\rangle
 +{\rm i} \left\langle \omega_{\alpha} \vert \left\lbrace
 \rho_{\alpha} - \widehat{\rho}_{\alpha}
  \right\rbrace \right\rangle
 \right).
\end{eqnarray}
Here  $\rho_{\alpha}({\mathbf r})$ is the collective variable
which describes the field of the number particle density of the $\alpha$th species.The functional integrals which enter the above equation can be given
a precise meaning  in the case where  the domain of volume $V$ occupied by particles  is a cube of side $L$ with periodic boundary conditions  which will be implicitly  assumed henceforth
\cite{caillol_patsahan_mryglod,Patsahan_Mryglod_Caillol-JPS}. This means that we restrict ourselves to the fields $\rho_{\alpha}({\mathbf r})$ and
$\omega_{\alpha}({\mathbf r})$ which can be  written as Fourier series.

Inserting equation (\ref{b}) in the definition (\ref{2.4}) of the
grand canonical partition function  one obtains an exact functional
representation
\begin{equation}
\Xi\left[\{ {\nu_{\alpha}} \}\right]=\int \mathcal{D} \rho
\mathcal{D} \omega\; \exp \left(- {\mathcal
H}[\{\nu_{\alpha}\},\{\rho_{\alpha},\omega_{\alpha}\}]\right) \;,
\label{VSS}
\end{equation}
where the action ${\mathcal H}[\{\nu_{\alpha}\},\{\rho_{\alpha},\omega_{\alpha}\}]$ of the CV field theory reads as
\begin{equation}
\label{actionCV}
\mathcal{H} \left[\{\nu_{\alpha}\}, \{\rho_{\alpha},
\omega_{\alpha}\} \right]=\frac{\beta}{2} \left\langle \rho_{\alpha}
\vert w_{\alpha\beta}\vert \rho_{\beta} \right\rangle  -{\rm i}
\left\langle \omega_{\alpha} \vert \rho_{\alpha}\right\rangle -
\ln\Xi_{\mathrm{RS}}\left[\{ \overline{\nu}_{\alpha} -{\rm i}
\omega_{\alpha}\} \right] \; .
\end{equation}
In (\ref{actionCV}) $\Xi_{\mathrm{RS}}[\{\overline{\nu}_{\alpha}-{\rm i}\omega_{\alpha}\}]$ is the grand canonical
partition function of a $m$-component RS defined as follows:
\begin{eqnarray}
\Xi_{\mathrm{RS}}[\{\nu_{\alpha}^{*}\}]&=&\sum_{N_{1}\geqslant  0}\frac{1}{N_{1}!}\sum_{N_{2}\geqslant  0}\frac{1}{N_{2}!}
\ldots\sum_{N_{m}\geqslant  0}\frac{1}{N_{m}!}\int({\rm d}\Gamma)
\exp\left(-\beta {\cal V}^{\mathrm{RS}}_{N_{1}\ldots N_{m}}+\langle\nu_{\alpha}^{*}\vert\widehat\rho_{\alpha}\rangle\right)
\label{2.11a}
\end{eqnarray}
with
\begin{equation}
\nu_{\alpha}^{*}({\mathbf r})=\overline{\nu}_{\alpha}(r)-{\rm
i}\omega_{\alpha}({\mathbf r}).
\label{nu_eff}
\end{equation}
Some comments are in order. It should be emphasized  that the description (\ref{VSS})--(\ref{nu_eff}) is based on the two
sets of variables $\{\rho_{\alpha}\}$
and $\{\omega_{\alpha}\}$ and  valid  for
repulsive, attractive as well as  arbitrary pair interactions.  One can distinguish the following alternative approaches to the
application of equations~(\ref{VSS})--(\ref{nu_eff}):
\begin{itemize}
\item Let  ${\mathbf{W}}$ denote the matrix of elements $w_{\alpha\beta}(\mathbf{r},\mathbf{r'})$.
If  ${\mathbf{W}}$ is the  positive-definite matrix, the integration over
$\{\rho_{\alpha}\}$ can be easily performed.
As a result, one arrives  at the same functional representation  as that obtained by means of the HS transformation.  Thus, we stress that the HS transformation has
a narrower region of applicability compared to the CV method.
\item
Let matrix ${\mathbf{W}}$  be  arbitrary. In this case, one can
start  with integration over $\{\omega_{\alpha}\}$. In general,
$\Xi_{\mathrm{RS}}[\{\nu_{\alpha}^{*}\}]$ cannot be calculated
exactly. In order to develop   a perturbation theory  we  present
the logarithm of the grand partition function  of the RS in the form
of a cumulant expansion
\begin{eqnarray}
\ln\Xi_{{\mathrm RS}}[\{\nu_{\alpha}^{*}\}]&=&\sum_{n\geqslant  0}\frac{(-{\rm
i})^{n}}{n!}\sum_{\alpha_{1},\ldots,\alpha_{n}} \int {\rm
d}1\ldots\int {\rm d}n\,{\mathfrak{M}}_{\alpha_{1}\ldots\alpha_{n}}(1,\ldots,n)
\omega_{\alpha_{1}}(1)\ldots\omega_{\alpha_{n}}(n),
\label{2.16}
\end{eqnarray}
where $i\equiv {\mathbf r}_{i}$ and ${\rm d}i\equiv{\rm d}{\mathbf
r}_{i}$. In (\ref{2.16}) the $n$th cumulant
${\mathfrak{M}}_{\alpha_{1}\ldots\alpha_{n}}(1,\ldots,n)$
is equal to the $n$-particle partial truncated (connected)
correlation function of RS at
$\nu_{\alpha}^{*}({\mathbf r})=\overline\nu_{\alpha}({\mathbf r})$. Substituting
(\ref{2.16}) in (\ref{VSS}) we obtain
\begin{eqnarray}
\Xi\left[\{\nu_{\alpha}\}\right]&=&\Xi_{{\mathrm
RS}}\left[\{\overline\nu_{\alpha}\}\right]\int \mathcal{D} \rho
\mathcal{D} \omega\;\exp\Bigg\{ -\frac{\beta}{2}
\left\langle \rho_{\alpha} \vert w_{\alpha\beta}\vert
\rho_{\beta} \right\rangle
\nonumber \\
&& +{\rm i} \left\langle \omega_{\alpha} \vert
\rho_{\alpha}\right\rangle +\sum_{n\geqslant  1}\frac{(-{\rm
i})^{n}}{n!}\sum_{\alpha_{1},\ldots,\alpha_{n}} \int {\rm
d}1\ldots\int {\rm d}n\,
{\mathfrak{M}}_{\alpha_{1}\ldots\alpha_{n}}(1,\ldots,n)
\omega_{\alpha_{1}}(1)\ldots\omega_{\alpha_{n}}(n)
\Bigg\}  \; . \qquad \label{Xi_full}
\end{eqnarray}
The calculation of correlation functions of RS,
${\mathfrak{M}}_{\alpha_{1}\ldots\alpha_{n}}(1,\ldots,n)$,  is a
separate task. If RS is a hard sphere mixture, one can use the
fundamental-measure theory \cite{Rosenfeld} and the Percus-Yevick or
Car\-na\-han-Starling approximations \cite{leb1,Mansoori:71} in the
non-uniform and uniform cases, respectively.

Equation~(\ref{Xi_full})  can be evaluated  in a systematic way using the Gaussian distribution as a basic one. In particular, using the Gaussian averages
one can develop a loop expansion of the grand partition function  as it was done recently
for a one-component fluid \cite{caillol_patsahan_mryglod}.

\item
Another way of integrating over $\{\omega_{\alpha}\}$  is to use  the steepest  descent method.  If RS is  a mixture of ideal gases,
the integration in (\ref{2.11a})  can be performed exactly.
We consider this special case in detail.
\end{itemize}
Let RS  be a $m$-component mixture of point particles that corresponds, in turn,  to the condition ${\cal V}^{RS}_{N_{1}\ldots
N_{m}}=0$ in equation (\ref{2.11a}). In this case the right hand
side of (\ref{2.11a}) can be easily calculated
\begin{eqnarray}
\Xi_{\mathrm{id}}[\{\nu_{\alpha}^{*}\}]&=&\sum_{N_{1}\geqslant  0}\frac{1}{N_{1}!}\sum_{N_{2}\geqslant  0}\frac{1}{N_{2}!}
\ldots\sum_{N_{m}\geqslant  0}\frac{1}{N_{m}!}\,\prod_{\alpha=1}^{m}\int\,{\rm d}\mathbf{r}_{1}^{\alpha}\ldots
{\rm d}\mathbf{r}_{N_{\alpha}}^{\alpha}
\exp\left(\sum_{i=1}^{N_{\alpha}}\,\nu_{\alpha}^{*}(\mathbf{r}_{i}^{\alpha})\right)
\nonumber \\
& =&\prod_{\alpha=1}^{m}\left[\sum_{N_{\alpha}\geqslant
0}\frac{1}{N_{\alpha}!}\left(\int \,{\rm
d}\mathbf{r}\exp(\nu_{\alpha}^{*}(\mathbf{r})
\right)^{N_{\alpha}}\right] = \exp\left[\sum_{\alpha}\int\, {\rm
d}\mathbf{r}\,\exp\left(\nu^{*}_{\alpha}(\mathbf{r})\right)\right].
\label{RS_ID}
\end{eqnarray}
Using (\ref{RS_ID})  we can rewrite the action (\ref{actionCV}) as
follows:
\begin{equation}
\label{actionCV_IDEL}
\mathcal{H} \left[\{\nu_{\alpha}\}, \{\rho_{\alpha},
\omega_{\alpha}\} \right]=\frac{\beta}{2} \left\langle \rho_{\alpha}
\vert w_{\alpha\beta}\vert \rho_{\beta} \right\rangle  -{\rm i}
\left\langle \omega_{\alpha} \vert \rho_{\alpha}\right\rangle -
\sum_{\alpha}\int\, {\rm d} {\mathbf r}\exp\left(\overline{\nu}_{\alpha}(\mathbf r) -{\rm i}
\omega_{\alpha}(\mathbf r)\right).
\end{equation}
In (\ref{actionCV_IDEL}) we take into account the
equation~(\ref{nu_eff}).  It is worth noting that in a uniform
case, the  action (\ref{actionCV_IDEL}) coincides with the corresponding
expression obtained in \cite{honopolski}.

In order to integrate in (\ref{VSS}) over the CV fields
$\omega_{\alpha}({\mathbf r})$ we use the steepest
descent method. In this case the equation
\begin{equation*}
 \left.\frac{\delta {\mathcal H}\left[\{\nu_{\alpha}\}, \{\rho_{\alpha},
\omega_{\alpha}\}
\right]}{\delta\omega_{\alpha}}\right\vert_{\omega_{\alpha}=\overline\omega_{\alpha}}=0
\end{equation*}
leads to the relation
\begin{equation}
\rho_{\alpha}(\mathbf{r})= \exp\left(\overline\nu_{\alpha}(\mathbf{r})-{\rm
i}\overline\omega_{\alpha}(\mathbf{r})\right). \label{rho_omega}
\end{equation}
Introducing the notation $\rho_{\alpha}^{0}=\exp(-\ln\Lambda_{\alpha}^{3}+\nu_{\alpha}^{s})$ we can rewrite (\ref{rho_omega}) as
\begin{equation}
 -{\rm i}\overline\omega_{\alpha}(\mathbf{r})=\ln\frac{\rho_{\alpha}(\mathbf{r})}{\rho_{\alpha}^{0}}-\beta\mu_{\alpha}+
\beta\psi_{\alpha}(\mathbf{r}).
\label{rho_omega1}
\end{equation}
Substituting (\ref{rho_omega1}) in equation  (\ref{actionCV_IDEL}) we finally get for $\Xi\left[\{ {\nu_{\alpha}} \}\right]$
\begin{equation}
\Xi\left[\{ {\nu_{\alpha}} \}\right]=\int \mathcal{D} \rho \; \exp
\left(- {\cal H}[\{ \nu_{\alpha}\},\{\rho_{\alpha}\}]\right)\;, \label{VSS_sad}
\end{equation}
where the action ${\cal H}[\{\rho_{\alpha}\}]$ has the form:
\begin{eqnarray}
&&{\cal H}[\{ \nu_{\alpha} \},\{\rho_{\alpha}\}]=\sum_{\alpha}\int\, {\rm d} {\mathbf
r}\,\rho_{\alpha}(\mathbf{r})\left[\ln\frac{\rho_{\alpha}(\mathbf{r})}{\rho_{\alpha}^{0}}-1\right]
+\frac{\beta}{2} \left\langle \rho_{\alpha}
\vert w_{\alpha\beta}\vert \rho_{\beta} \right\rangle
-\left\langle\rho_{\alpha}\vert\beta\mu_{\alpha}\right\rangle+\left\langle\rho_{\alpha}\vert\beta\psi_{\alpha}\right\rangle.
\label{action_FT}
\end{eqnarray}
It is remarkable that  the contribution  from quadratic fluctuations
of $\Delta\omega_{\alpha}=\omega_{\alpha}- \overline\omega_{\alpha}$
vanishes in the thermodynamic limit \cite{Frusawa_Hayakawa}.
Therefore, the functional representation of the grand partition
functional given by (\ref{VSS_sad})--(\ref{action_FT}) is exact for
the model with the interaction potential (\ref{2.1})--(\ref{split})
under condition $v_{\alpha\beta}^{0}(r)=~0$.  This contradicts the
assumption made in \cite{Evans_2} about a purely phenomenological character of  functional
Hamiltonians which are employed in the field-theoretical formalism.

For a one-component case, equations
(\ref{VSS_sad})--(\ref{action_FT}) coincide   with the  functional
integral representation obtained in
\cite{Frusawa_Hayakawa,Woo_Song,diCaprio_Badiali:08}.

Based on (\ref{action_FT}) one can formulate  the MF theory. To this
end, from the stationary-point condition
\begin{eqnarray*}
\left.\frac{\partial{\cal
H}[\{ \nu_{\alpha} \},\{\rho_{\alpha}\}]}{\partial\rho_{\alpha}}\right\vert_{\rho_{\alpha}=\bar\rho_{\alpha}}=0
\end{eqnarray*}
we obtain the following set of equations for the MF density
\begin{eqnarray}
\ln \bar\rho_{\alpha}(r)=\overline\nu_{\alpha}(r)-\beta\sum_{\beta}\int
{\rm d} \mathbf{r'}
 w_{\alpha\beta}(\mathbf{r},\mathbf{r'})\bar\rho_{\beta}(\mathbf{r'}).
\label{station_point}
\end{eqnarray}
Then, from (\ref{VSS_sad})--(\ref{action_FT}), the MF grand potential
reads
\begin{eqnarray*}
\ln \Xi_{\mathrm{MF}}\left[\{ {\nu_{\alpha}}
\}\right]=\sum_{\alpha}\int\,{\rm d}\mathbf{r}
\bar\rho_{\alpha}(\mathbf{r})+\frac{\beta}{2} \left\langle
\bar\rho_{\alpha} \vert w_{\alpha\beta}\vert \bar\rho_{\beta}
\right\rangle.
\end{eqnarray*}
The  Helmholtz free energy of the system defined as the Legendre
transform
\begin{equation*}
F\left[\{\rho_{\alpha}\}\right] =\left\langle
\rho_{\alpha} \vert \mu_{\alpha}\right\rangle -\beta^{-1}\ln
\Xi\left[\{\nu_{\alpha}\}\right],
\end{equation*}
has the following form in the MF approximation
\begin{eqnarray*}
F_{\mathrm{MF}}\left[\{\rho_{\alpha}\} \right]& =
& \beta^{-1}\sum_{\alpha}\int\,{\rm d}\mathbf{r} \rho_{\alpha}(\mathbf{r})
\left[\ln\left(\frac{\rho_{\alpha}(\mathbf{r})}{\rho_{\alpha}^{0}}\right)
-1\right]+\frac{1}{2} \left\langle \rho_{\alpha}
\vert w_{\alpha\beta}\vert \rho_{\beta} \right\rangle+\left\langle\rho_{\alpha}\vert\psi_{\alpha}\right\rangle.
\end{eqnarray*}
Introducing an intrinsic free energy $\cal{F}$
\cite{hansen_mcdonald} by the relation ${\cal{F}}=F-\left\langle
\rho_{\alpha} \vert \psi_{\alpha}\right\rangle$ one can present the
MF intrinsic free energy of the system  as follows:
\begin{eqnarray}
\label{intrinsic_energy}
{\cal{F}}_{\mathrm{MF}}\left[\{\rho_{\alpha}\}\right]& =&
\beta^{-1}\sum_{\alpha}\int\,{\rm d}\mathbf{r} \rho_{\alpha}(\mathbf{r})
\left[\ln\left(\frac{\rho_{\alpha}(\mathbf{r})}{\rho_{\alpha}^{0}}\right)
-1\right]+\frac{1}{2} \left\langle \rho_{\alpha}
\vert w_{\alpha\beta}\vert \rho_{\beta} \right\rangle.
\end{eqnarray}
 Note that ${\cal F}_{\mathrm{MF}}\left[\{\rho_{\alpha}\}
\right]$ is the functional of the  local density
$\rho_{\alpha}(\mathbf{r})$ and has no explicit dependence on the
external potential $\psi_{\alpha}(\mathbf{r})$. Functional
(\ref{intrinsic_energy}) has the form of the approximate free energy
functional used in the DFT when the  excess free energy functional
arising from the interactions is treated at the MF level. In this
case the variational principle
\[
\frac{\delta }{\delta\rho_{\alpha}(\mathbf{r})}\Big[{\cal
F}\left[\{\rho_{\alpha}\} \right]-\left\langle \rho_{\alpha} \vert
(\mu_{\alpha}-\psi_{\alpha})\right\rangle\Big]=0
\]
leads to the equation (\ref{station_point}) and, in turn, to  a
coincidence of the MF density $\bar\rho_{\alpha}(\mathbf{r})$ with
the equilibrium local density.

Summarizing, we have derived  the exact field theoretical
representation (equations (\ref{VSS_sad})--(\ref{action_FT})) for the
$m$-component inhomogeneous  system that does not include the hard
sphere interaction. Within the framework of the MF formulation of
the theory we have found the  functional of the free energy which is
analogous to that used in the MF DFT.

It is worth noting that the long-range interactions within the
framework of the DFT are usually treated within the MF
approximation. By contrast, the CVs based theory enables one to
develop the perturbation scheme in order to take into account the
correlation effects. Moreover, the latter theory can be applied to
the  systems with more complicated interactions   than those
considered in this work.

We  have also demonstrated that the density functional integral
formulation derived in \cite{Frusawa_Hayakawa,Woo_Song,diCaprio_Badiali:08,diCaprio_Holovko_Badiali:03}
can be obtained from the exact CVs functional representation  in
some special case.

We note that in general the integration over $\{ \omega_{\alpha} \}$
can be performed exactly in thermodynamic limit if the RS describes
noninteracting particles. When short-range interactions are included
into the RS the method of steepest descent will produce additional
contributions describing the correlations between particles in the
RS. In particular, this is the case of a hard sphere RS that will be
considered elsewhere. However, even for RSs of noninteracting
particles there are several models where the method developed above
can be very useful. One of such examples is connected with the
statistical field theory of anisotropic fluids where the RS may be
considered as the system of noninteracting rigid rotators (see, for
instance,~\cite{Holovko_diCaprio_Kravtsiv:11}).


%
%

 \ukrainianpart
 \title{Метод колективних змінних: зв'язок з теорією функціоналу густини}
 \author{О. Пацаган, І. Мриглод}
 \address{Інститут фізики конденсованих систем НАН України, вул. Свєнціцького, 1, 79011 м. Львів, Україна}
%
 \makeukrtitle
 \begin{abstract}
Недавно, використовуючи метод колективних змінних, було сформульовано
статистико-польову теорію для багатокомпонентних неоднорідних систем
[O. Patsahan, I. Mryglod, J.-M. Caillol, Journal of Physical
Studies, 2007, {\textbf {11}}, 133]. В цьому повідомленні ми
встановлюємо зв'язок між цим підходом і класичною теорію функціоналу
густини для неоднорідних плинів.

 \end{abstract}


\begin{thebibliography}{99}
\bibitem{stratonovich} Stratonovich R.L., Sov. Phys. Solid State, 1958, {\textbf 2}, 1824.
%
\bibitem{hubbard} Hubbard J.,  Phys. Rev. Lett., 1959, {\textbf 2}, 77; \doi{10.1103/PhysRevLett.3.77}.
%
\bibitem{zubar}   Zubarev D.N.,  Dokl. Acad. Nauk SSSR, 1954, {\textbf{95}}, 757 (in Russian).
%
\bibitem{jukh} Yukhnovsky I.R., Zh. Eksp. Ter. Fiz., 1958, {\textbf{34}}, 379 (in Russian).
%
\bibitem{bohm}  Bohm D.,  Pines D., Phys. Rev., 1951, {\textbf{82}}, 625; \doi{10.1103/PhysRev.82.625}.
%
\bibitem{yuk}  Yukhnovskii I.R., Phase Transitions of the Second Order:
Collective Variables Method. World Scientific, Singapore, 1987.
%
\bibitem{yuk_nuovo_cimento} Yukhnovs'kii I.R., Rivista del Nuovo Cimento, 1989, {\bf 12}, No.~1, 1; \doi{10.1007/BF02740597}.

\bibitem{yuk2}  Yukhnovskii I.R., Proceedings of the Steklov Institute of Mathematics, 1992, {\bf 2}, 223.
%
\bibitem{patyuk3}  Patsagan O.V.,  Yukhnovskii I.R., Teor. Mat. Fiz., 1990, {\textbf{83}}, 72 (in Russian).
%
\bibitem{patyuk4}  Yukhnovskii I.R.,  Patsahan O.V., J. Stat. Phys., 1995, {\textbf{81}}, 647; \doi{10.1007/BF02179251}.
%
\bibitem{Yuk-Hol} Yukhnovskii I. R.,   Holovko M.F., Statistical Theory of Classical Equilibrium Systems. Naukova Dumka, Kiev, 1980 (in Russian).
%
\bibitem{hansen_mcdonald}   Hansen J.P.,   McDonald I.R., Theory of simple liquids. Academic Press, 1986.
%
%
\bibitem{Cai-Mol}  Caillol J.-M., Mol. Phys., 2003, {\textbf{101}}, 1617; \doi{10.1080/0026897031000068488}.
%
\bibitem{Cai-JSP}  Caillol J.-M., J. Stat. Phys., 2004, {\textbf{115}}, 1461; \doi{10.1023/B:JOSS.0000028066.25728.cf}.
%
\bibitem{caillol_patsahan_mryglod}  Caillol J.-M.,  Patsahan O., Mryglod I., Physica A, 2006, {\textbf{368}}, 326; \doi{10.1016/j.physa.2005.11.010}.
%
\bibitem{patsahan_mryglod_CM}  Patsahan O., Mryglod I., Condens. Matter Phys., 2006, {\textbf{9}}, 659.

%
\bibitem{Evans}  Evans R., Adv. Phys., 1979, {\textbf{28}}, No.~2,  143; \doi{10.1080/00018737900101365}.

%
\bibitem{Singh} Singh Y., Phys. Rep., 1991, {\textbf{207}}, 351; \doi{10.1016/0370-1573(91)90097-6}.
%
\bibitem{Patsahan_Mryglod_Caillol-JPS}  Patsahan O., Mryglod I., Caillol J.-M.,
Journal of Physical Studies, 2007, {\textbf {11}}, No.~2, 133.
%
\bibitem{Rosenfeld} Rosenfeld~Y., Phys. Rev. Lett., 1989, \textbf{63}, 980; \doi{10.1103/PhysRevLett.63.980}.
%
\bibitem{leb1}
 Lebowitz~J.L., Phys. Rev., 1964, {\textbf{133}}, 895; \doi{10.1103/PhysRev.133.A895}.
%
\bibitem{Mansoori:71} Mansoori G.A., Carnahan N.F., Starling K.E.,  Leland T.W., J. Chem. Phys., 1971,  {\textbf{54}}, 1523; \doi{10.1063/1.1675048}.
%
\bibitem{honopolski}  Yukhnovskii I.R., Honopolskii O.L., Preprint of the Institute
for Theoretical Physics, ITP--74--93P, Kiev, 1974, (in Russian).

%
\bibitem{Frusawa_Hayakawa} Frusawa  H., Hayakawa R., Phys. Rev. E, 1999, {\textbf 60}, R5048; \doi{10.1103/PhysRevE.60.R5048}.
%
%
\bibitem{Evans_2}  Evans R., Mol. Phys., 1981, {\textbf 42},
1169; \doi{10.1080/00268978100100881}.
%
\bibitem{Woo_Song}  Woo H.-J., Song X., J. Chem. Phys., 2001, {\bf 114},
5637; \doi{10.1063/1.1353553}.
%

%
\bibitem{diCaprio_Badiali:08} di~Caprio D., Badiali J.P., J. Phys. A: Math. Theor., 2008, {\textbf{41}}, 125401; \doi{10.1088/1751-8113/41/12/125401}.
%
\bibitem{diCaprio_Holovko_Badiali:03} di~Caprio D., Holovko M.F., Badiali J.P., Condens. Matter Phys.,
2003, {\textbf{6}}, 693.
%
\bibitem{Holovko_diCaprio_Kravtsiv:11}Holovko M., di Caprio D., Kravtsiv I., Condens. Matter Phys.,
2011, {\textbf{14}}, 33605; \doi{10.5488/CMP.14.33605}.
\end{thebibliography}
\end{document}